\documentclass[sigconf, 9pt]{acmart}
\AtBeginDocument{%
  }

\usepackage{balance}
\usepackage[linesnumbered,ruled,vlined]{algorithm2e} % Options for styling
\begin{document}

%\title{On-premises Database Search for Regulated Industries or Deep Context Database Search for Regulated Industries}
% \title{Deep Context Database Search for Regulated Industries}
% \title{Structured Data Search for Credit Unions- Driven by Automated Context Graphs}
\title{Tursio for Credit Unions: Structured Data Search with Automated Context Graphs}
% \title{Structured Data Search with Automated Context Graph}

\author{Shivani Tripathi, Ravi Shetye,  Shi Qiao, Alekh Jindal}

\email{research@tursio.ai}
\affiliation{%
  \vspace{0.2cm}
  \institution{Tursio}
  \city{Bellevue}
  \country{USA}
  \vspace{0.2cm}
}

% \affiliation{%

% \affiliation{%

% \affiliation{%

% \affiliation{%

% \affiliation{%

% \affiliation{%

% \affiliation{%


\begin{abstract}
Extracting actionable insights from structured databases in regulated industries is often hindered by complex schemas, legacy systems, and stringent data governance requirements.
We present Tursio, a secure, on-premises database search platform that enables business users to query enterprise databases using natural language.
Tursio automatically infers a \emph{context graph}---a schema-level metadata structure that captures join paths, column semantics, and domain annotations---and uses it to systematically generate accurate query plans through LLM-assisted compilation, grounding, and rewriting.
Unlike existing AI/BI tools that require extensive manual context curation, Tursio automates this end-to-end and deploys entirely on-premises.
We demonstrate Tursio through realistic scenarios in the credit union domain, and discuss its applicability to other regulated settings.
A demonstration video is available.\footnote{\url{https://blog.tursio.ai/cu_demo_audio_video_gq-1-mp4/}}
\end{abstract}

\maketitle

% Suggested flow:
% 1. Credit unions and Symitar background
% 2. Current search approaches involve manual context
% 3. Tursio automates that context on-premises:
%   - inferring context graph
%   - context-aware query planning
%   - one result
% 4. Demonstration scenarios
% 5. Early feedback and conclusion

\section{Introduction}

Credit unions (CUs) are not-for-profit financial cooperatives owned and operated by their members. They prioritize member service over profit maximization, often resulting in better rates and lower fees compared to traditional banks. As of 2025, there are more than $4{,}500$ CUs in the United States serving over $144$ million members~\cite{CUFinancialSummary}. Globally, this number goes up to $85{,}000$ CUs serving $274$ million members worldwide~\cite{WOCCU2023Stats}. Despite their reach, CUs have limited resources and face increasing pressure to keep pace with the larger banks. Therefore, understanding member lives better and delivering services personalized to their financial dreams is crucial for credit unions' survival and growth.

Unfortunately, getting member insights from operational data is challenging for credit unions. The most widely adopted core banking platform in the credit union industry, Symitar from Jack Henry~\cite{Shalabi2025Top20Cores}, was designed in the late 1970s to cater the operational needs. It has highly normalized (3NF) hierarchical schema, centered around Account with many related tables, which is tedious for analytics and reporting. Consequently, various stakeholders send data requests to report writers or the IT team, who manually write scripts to extract the required information. This process is time-consuming, taking hours to days, leading to delays in decision-making and missed opportunities for member engagement.

% Credit unions rely heavily on specialized core banking systems such as Symitar to manage their operations.
% Symitar, is now the industry standard and powers hundreds of credit unions nationwide \cite{Shalabi2025Top20Cores}.
% Symitar data in highly normalized (typically 3NF), organized as hierarchical account-centric schema with the account at the top level and all related tables tied to account IDs.
% While this structure is optimized for operational use cases, it poses challenges for member-centric analyses.
% It requires to evolve the schema to be member-centric by carefully de-normalizing tables while ensuring aggregates are not duplicated.

\begin{figure}[!t]
  \vspace{-0.2cm}
  \includegraphics[width=0.475\textwidth]{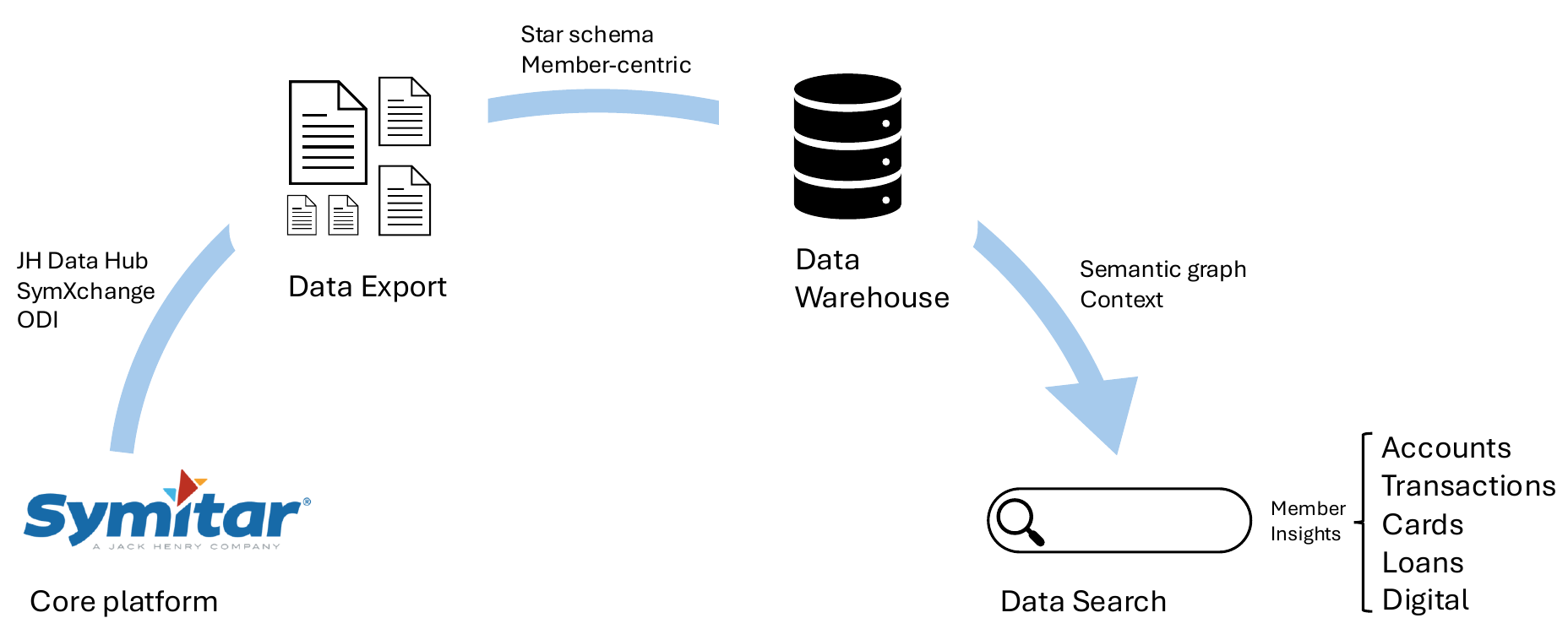}
  \vspace{-0.4cm}
  \caption{The journey of Symitar core banking data.}
  \Description{The journey of Symitar core banking data.}
  \label{fig:Symitar-journey}
  \vspace{-0.4cm}
\end{figure}

Current approach to address the above challenges is to modernize Symitar data by moving it into a relational or data warehousing platform and preparing it for analytics and reporting. Figure~\ref{fig:Symitar-journey} shows the journey of Symitar data from core platform to a data warehouse. Raw data fields in the Symitar system can be accessed through Operational Data Integration (ODI), the SymXchange API, or the Jack Henry Data Hub. ODI allows data to be exported as CSV files, while the SymXchange API enables third-party vendors to define data services that can be consumed by ETL processes. The Data Hub is a more modern solution that leverages Google Cloud Platform (GCP) to deliver data in standardized formats and integrate efficiently with data warehouses.

Despite the modernization efforts, stakeholders in a CU are still not self-served and they rely on experts for pulling data out. In fact, searching for information is a broader productivity problem; recent Gartner study reveals that $47\%$ of employees struggle to find the information they need to perform their jobs effectively~\cite{gartner-study}, while Microsoft report says $62\%$ of digital workers spend too much time hunting for information~\cite{MicrosoftSearch2023}.
Therefore, the journey of Symitar data is incomplete until it becomes easily searchable by a broader set of stakeholders, i.e., taking the operational data from the Symitar data goldmine all the way to refined insights.

In this paper, we present how Tursio supports CUs in that final step --- enabling structured data search for faster member insights.
Tursio's key contribution is automating the construction of a \emph{context graph}---a schema-level metadata structure that integrates inferred join paths (via inclusion dependency discovery~\cite{Rostin2009AML}), column classifications, and domain annotations---and coupling it with a multi-stage, LLM-assisted query planning pipeline that compiles natural language into grounded, rewritten SQL.
While the individual techniques (schema discovery~\cite{10.1145/588111.588141}, NL-to-SQL~\cite{bird_bench, text2sql_survey}, enterprise metadata management~\cite{sequeda2022enterprise}) are well-studied, Tursio's contribution lies in their end-to-end integration into a single system that requires no manual context curation and operates entirely on-premises---a hard requirement in regulated industries.
Although we focus on credit unions here, the approach generalizes to any domain with complex relational schemas; we have deployed Tursio in healthcare and insurance settings as well~\cite{clinical_search_tursio}.
Below we discuss the background and challenges in more detail, followed by Tursio's architecture and demonstration scenarios.
\vspace{-0.3cm}

\section{Background: Manual Context}

A new breed of AI/BI tools, such as Snowflake Cortex, Databricks Genie, ThoughtSpot Spotter, and Microsoft Fabric Copilot, provide mechanisms to manually build the relevant context, which can be later used for querying. Primarily, this includes three steps, namely building a \textbf{knowledge store}, compiling \textbf{sample questions}, and providing any \textbf{custom instructions}.

% \vspace{0.1cm}
{\bf Knowledge store.}
Modern AI/BI tools require data teams to manually configure a knowledge store to provide context for search.\
This involves: (1) enriching metadata with business definitions and synonyms, (2) customizing datasets for specific use cases, (3) sampling values to improve query matching, (4) defining join relationships for accurate queries, and (5) specifying business logic via SQL expressions.
For example, Databricks Genie and Snowflake Cortex both require mapping business concepts to physical tables, defining relationships and metrics, and creating semantic views to bridge user intent with underlying data structures.

% \vspace{0.1cm}
{\bf Sample questions.}
Once the knowledge store is configured, users need to provide sample questions paired with their corresponding SQL queries to enhance accuracy.
For instance, Databricks Genie and Snowflake Cortex both support curated repositories of example queries—Genie allows up to 100 instructions, while Cortex uses a Verified Query Repository (VQR) defined in its semantic model.
These examples guide the system in generating correct SQL for common or similar questions, ensuring consistency and reliability by referencing typical user patterns.

% \vspace{0.1cm}
{\bf Custom instructions.}
Custom instructions allow users to specify business-specific logic, terminology, or formatting rules that guide query generation and interpretation.
These instructions can define concepts (e.g., ``financial year''), enforce default filters, or clarify domain terminology.
They apply globally unless scoped to specific queries or tables, ensuring that guidance is targeted and improves SQL accuracy without cluttering the general context.

% \vspace{0.1cm}
Manually building the context as described above is challenging due to the effort involved.
Maintaining knowledge stores and sample questions is a continuous and tedious process. Furthermore, most AI/BI tools are cloud-based, exposing sensitive metadata and prompts to external environments,
which is an issue for regulated industries like CUs that require strict data control.
Finally, AI/BI tools are tied to the data platform with usage-based pricing models, adding unpredictability for limited resource organizations like CUs.

% \vspace{0.1cm}
{\bf Related work.}
Tursio builds on several lines of research.
Schema discovery and join inference have been studied extensively, from inclusion dependency detection~\cite{Rostin2009AML} to broader data integration~\cite{10.1145/588111.588141, halevy2006data}.
Enterprise knowledge graphs~\cite{sequeda2022enterprise} provide rich ontological representations of business data; Tursio's context graph is deliberately lighter-weight, capturing only the metadata needed for query planning (joins, column types, annotations) rather than full ontological modeling.
The NL-to-SQL problem has seen significant progress with benchmarks like BIRD~\cite{bird_bench} and Spider~\cite{spider_bench}, and LLM-based approaches~\cite{text2sql_survey, din_sql}; Tursio's multi-stage pipeline decomposes the problem into compilation, grounding, and rewriting steps rather than single-shot SQL generation.
Unlike prior systems that assume either a curated semantic layer or a clean star schema, Tursio infers context graph directly on complex normalized schemas.

\section{Tursio: Automated Context}

Tursio automates the context-building process, thereby reducing the manual effort for the CUs.
% Tursio is designed to facilitate context-aware search over structured data using automated context graphs.
% We also expose a simplified search interface that allows user to interact with the database in natural language rather than generating SQL queries that are hard to verify.
It first infers a context graph from the connected database and then uses it to generate query plans from natural language queries. The search interface supports result exploration through tabular views, visualizations, natural language explanations, query suggestions, and saved/shared queries.
We have packaged the product for on-premises deployment as a standalone Docker container or within Kubernetes clusters. Tursio is also available on Microsoft Azure Marketplace for deployment on private VMs. Additionally, we provide privacy, security, and quality knobs for the administrators to control the system behavior.

% As soon as Tursio connects to the database, we automatically build context graph that captures intricate relationships of the underlying database. The knowledge graph captures the semantics of the data, like name simplification, join inference, dimensions and measures, table and column descriptions, ontologies, data and value types, value aspects, valid aggregations, default measures, aliases, personally identifiable information (PII), sample values, and others. 

% This well-defined semantics of the data is later used to contextualize the user intent in the natual language query and guide the generation of accurate and efficient query plans.
% Tursio generates a query plan by parsing query fragments into meaningful operators, grounding the parsed operators into valid ones using metadata, generating a tree of well-formed operators, and further rewriting the query plan to be more expressive in answering the user questions as closely as possible.
% Finally, Tursio provides a rich search interface for business users to interact with the results.

% Figure~\ref{fig:tursio-steps} illustrates the three-step process of deployment, context-building, and search in Tursio.

% \begin{figure}[!t]
%   \includegraphics[width=0.475\textwidth]{pdf/tursio-steps.pdf}
%   \caption{The three-step process in Tursio.}
%   \Description{Three-step process in Tursio.}
%   \label{fig:tursio-steps}
%   \vspace{-0.4cm}
% \end{figure}

\subsection{Context Graph}

Tursio automatically infers a \emph{context graph} over the connected database.
Formally, a context graph $G = (T, J, A)$ consists of a set of tables $T$ as nodes, a set of inferred join edges $J$ (each with a join condition and confidence score), and a set of per-column annotations $A$ (including dimension/measure classification, descriptions, aliases, PII flags, and sample values).
Unlike enterprise knowledge graphs~\cite{sequeda2022enterprise} that model domain ontologies, a context graph captures only the schema-level metadata needed for query planning.
Figure~\ref{fig:semantic-model} shows an example context graph inferred by Tursio over the sample credit union database.
Once users connect a database and select the tables to include, we profile it by collecting fixed-sized samples and computing statistics.
This is leveraged to build the context graph via: (1)~Schema Inference, and (2)~Semantic Enrichment.

% \vspace{0.1cm}
{\bf Schema inference.}
Tursio identifies key/foreign-key relationships between tables using an inclusion dependency discovery algorithm~\cite{Rostin2009AML}.
We first detect the primary keys using a combination of statistics and then identify candidates for inclusion dependencies by assigning a weighted score of schema similarity and value inclusions.
We further prune these candidates using heuristics and validate them using LLMs to generate high-fidelity join conditions. Overall, this is a scalable join inference approach, and we have successfully onboarded schemas ranging from 10s to 100s of tables.

% \vspace{0.1cm}
{\bf Semantic enrichment.}
Tursio classifies columns as dimensions or measures using schema information, statistics, and LLMs (e.g., ``customer name'' is a dimension while ``order amount'' is a measure).
Furthermore, we use LLM to expand abbreviated columns names to human-friendly names, generate descriptions for columns and tables and use them as context to disambiguate the user intent and improve response quality. 
Tursio also detects and excludes personally identifiable information (PII) during query planning to avoid accidental exposure of sensitive data. 
Tursio supports custom measures, where obvious ones can be generated automatically, or users can provide custom SQL expressions or import it from existing BI tools that will be inferred when compiling natural language queries.
Finally, Tursio generates compact aliases for tables and columns names that are too verbose for better grounding.

Overall, the context graph captures a holistic understanding of the database, while still allowing users to refine it further.

\begin{figure}[!t]
  \includegraphics[width=0.475\textwidth]{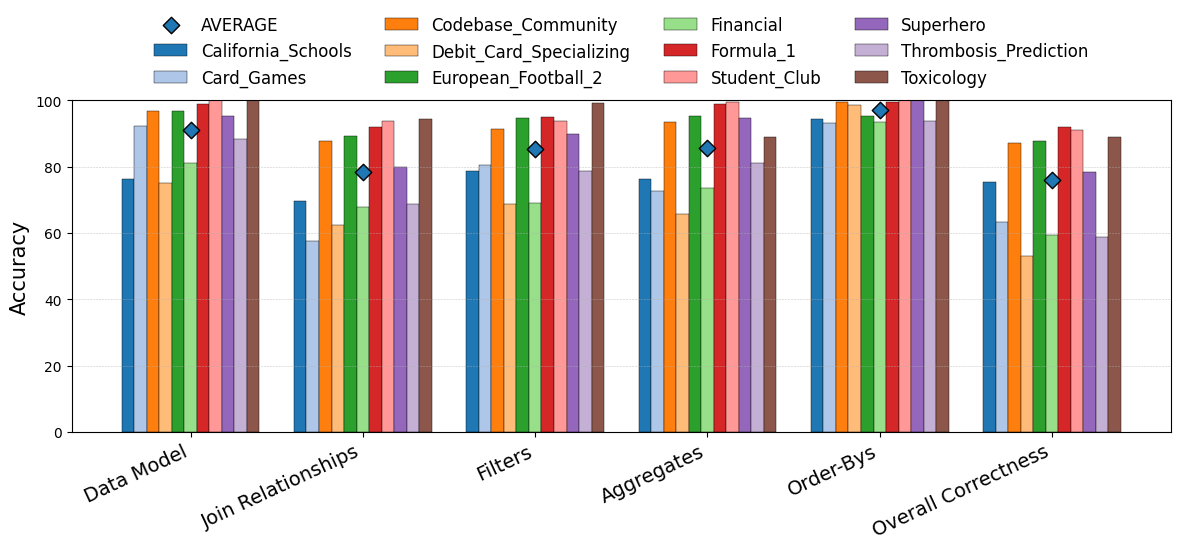}
  \vspace{-0.4cm}
  \caption{Evaluating SQL structural accuracy (BIRD)}
  \Description{SQL structural accuracy (BIRD)}
  \label{fig:sql-structural-accuracy}
  \vspace{-0.4cm}
\end{figure}

\subsection{Context-Aware Query Planning}

Tursio processes natural language queries by first contextualizing the user intent to relevant portions of the context graph.
This involves two main steps: (i)~identifying the input tables (data models) needed to answer the query, and (ii)~identifying the operations involved to answer the query.
We first determine the data models by mapping keywords to tables using a hash-based predictor, supplemented by semantic search and join graph traversal with LLM adjudication to resolve ambiguities.
We then identify the necessary operations (e.g., projection, filtering, aggregation) by leveraging pre-generated sample questions and LLMs to interpret user intent against the context graph.

Once identified the input tables and the operations, Tursio generates query plans from natural language queries using a systematic approach.
The key idea is to break down the query planning into several LLM-assisted steps to handle the complexity of natural language. The main steps are as follows:
(1)~parsing queries into operators using LLMs and ANTLR-based parsers,
(2)~grounding these operators to valid schema elements via the context graph and multiple matching techniques,
(3)~composing a relational operator tree,
(4)~applying rule-based transformations for correctness and security,
(5)~generating executable SQL, and
(6)~optionally rewriting queries with LLMs for advanced SQL constructs.
This stepwise process incrementally reduces ambiguity and ensures accurate, secure, and expressive query generation.

\begin{figure}[!t]
  \includegraphics[width=0.475\textwidth, height=0.23\textwidth]{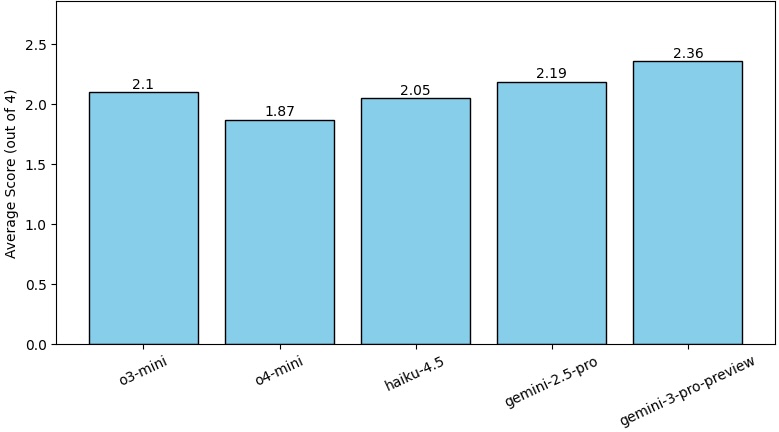}
  \vspace{-0.4cm}
  \caption{Evaluating text response quality.}
  \Description{Evaluating text response quality.}
  \label{fig:response_quality}
  \vspace{-0.5cm}
\end{figure}

{\bf Evaluation.} We evaluated Tursio on the BIRD-DEV benchmark~\cite{bird_bench} for structural accuracy of generated SQL (i.e., whether the predicted SQL uses the correct tables, joins, columns, filters, and aggregations compared to the reference SQL), using an LLM-based grader validated against human judgments.
Figure~\ref{fig:sql-structural-accuracy} shows component-wise results; predicted and reference SQLs align well across most dimensions.
We also evaluated response quality on an internal production workload (scale of 1--4, 1 = Excellent) using an ensemble of LLM graders.
Figure~\ref{fig:response_quality} shows results across models, with a best score of 1.8.
In terms of latency, the median end-to-end response time is $8.2$s on the production workload, with LLM inference accounting for $\sim$70\% of total time. Context graph construction is a one-time cost of $5$--$30$ minutes depending on schema size.
% Our full paper is under submission.

% We have described the Tursio's architecture and query processing in detail in our prior work (currently under submission).

\section{Demonstration Scenarios}

\begin{figure}[!t]
  \includegraphics[width=0.475\textwidth]{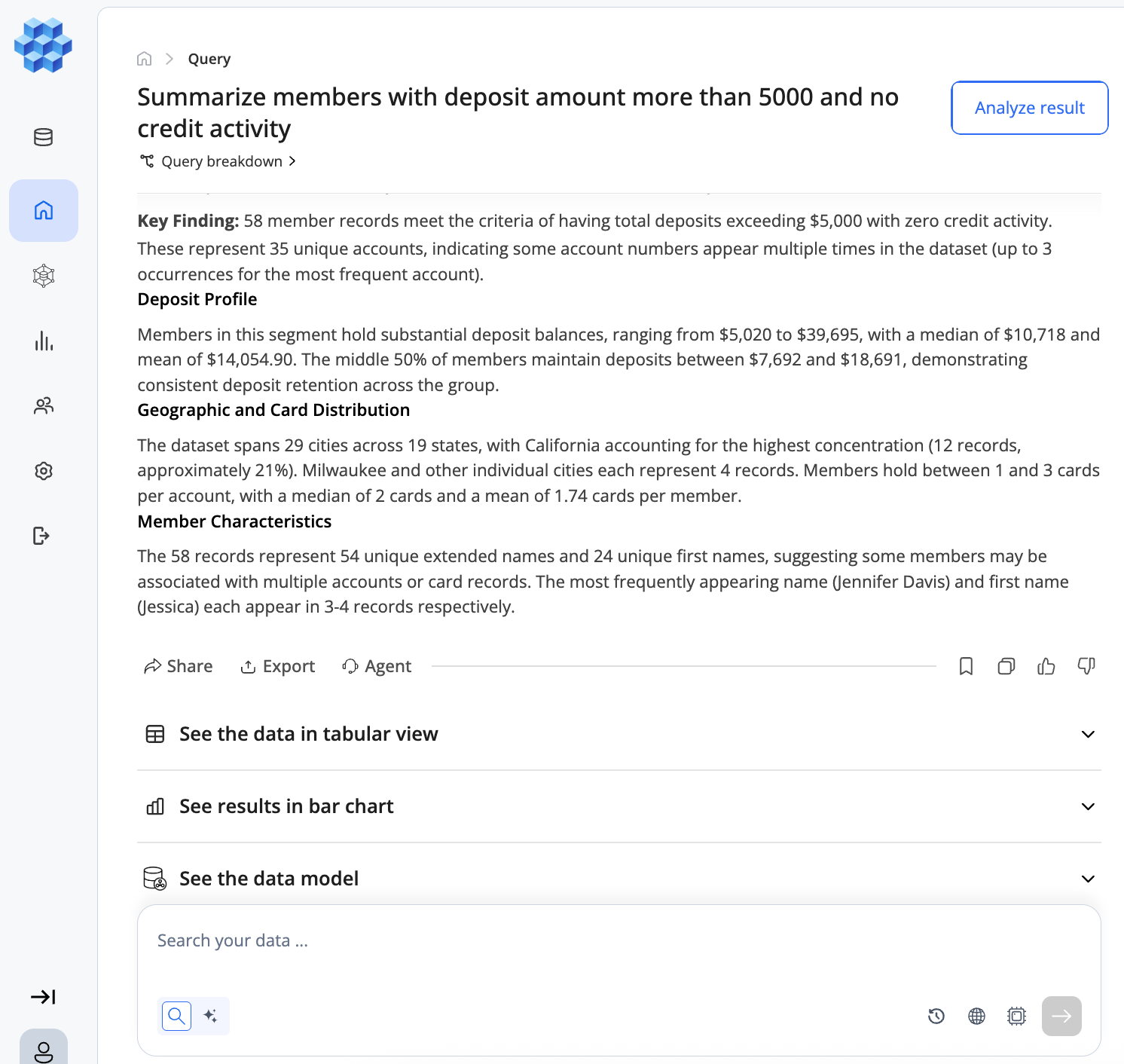}
  \caption{Tursio search interface.}
  \Description{Query Answer}
  \label{fig:query-summary}
  \vspace{-0.5cm}
\end{figure}

Let us now dive into the demonstration scenarios, derived from day-to-day operational needs of CUs, powered by the Tursio search.
%  interface that provides a rich set of features for business users to interact with the results. 
Figure~\ref{fig:query-summary} shows the interface with a typical CU question.
% We demonstrate Tursio's capabilities through a set of realistic scenarios derived from day-to-day operational needs of credit unions.
We will invite the attendees to play with Tursio's search interface using a sample CU database with $5$ core tables and synthetic operational data of over $1{,}000$ accounts.
% The users will be able to use either one of the following scenarios or bring their own scenarios to interactively explore Tursio's features.

% \vspace{0.1cm}
% \textbf{Member-centric Questions Over Account-centric Schema.}
\textbf{Accurate member insights.}
Given that traditional Symitar schema is account-centric, CUs often struggle to extract member-centric insights. 
% Member-centric questions over account-centric schemas are critical because they reflect how credit unions serve and understand their members, rather than just managing accounts.
% Traditional schemas are optimized for transactions and operations, not for holistic member insights.
In this scenario, we demonstrate how Tursio automatically joins across the account-centric schema to answer member-centric questions, without requiring users to understand the underlying table relationships.
Credit unions may ask questions centered around members rather than querying accounts directly, such as {\it ``Which members have closed accounts in the last quarter?''}, {\it ``List members with delinquent loans exceeding $\$5,000$.''}, etc.
Tursio ensures to de-normalize the schemas and handle symmetric aggregates correctly to avoid double counting~\cite{google-looker-symmetric-aggregates}.

\begin{figure}[!t]
  \includegraphics[width=0.475\textwidth]{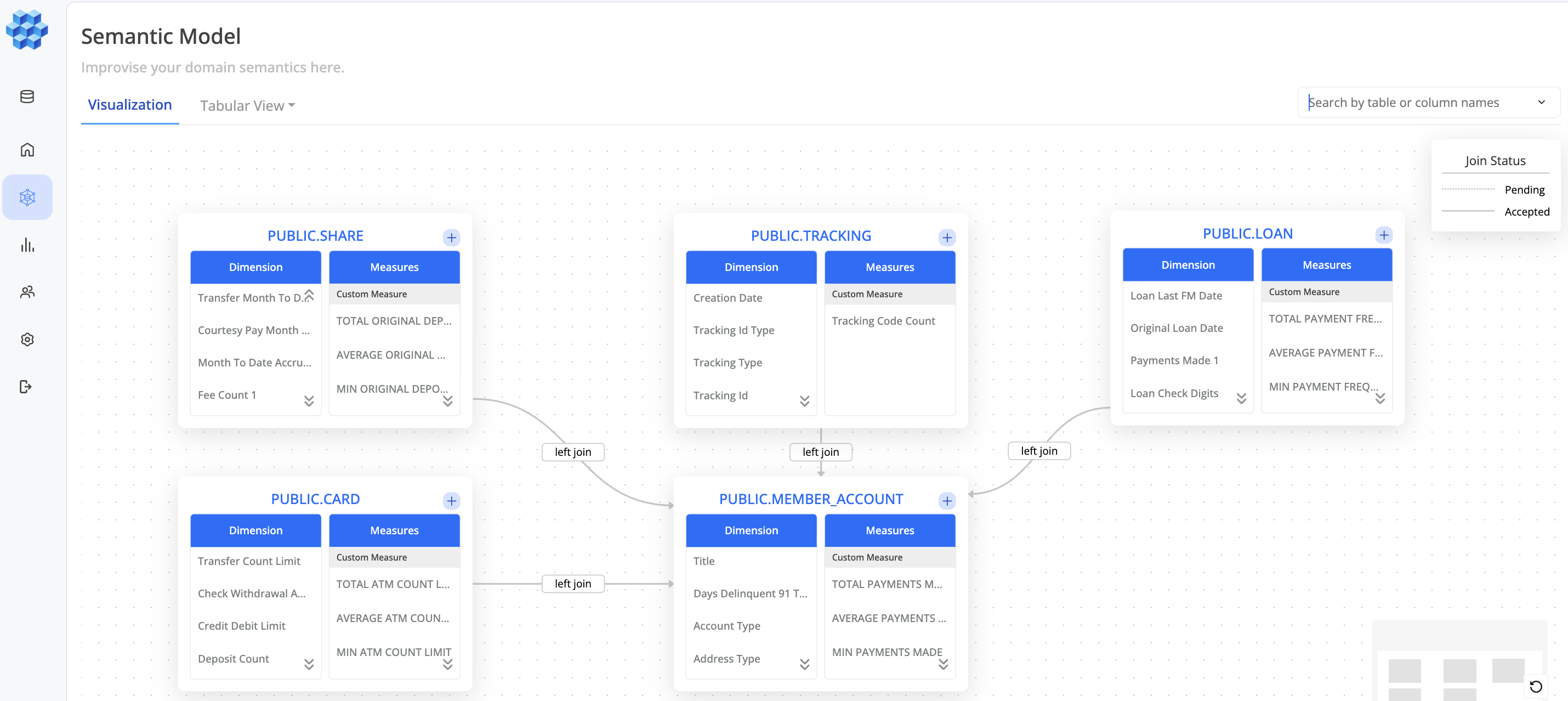}
  \caption{Context graph for sample CU database, including tables, inferred joins, and associated annotations.}
  \Description{Context graph for CU database}
  \label{fig:semantic-model}
  \vspace{-0.5cm}
\end{figure}

We further help user visualize and edit the context graph (Figure \ref{fig:semantic-model}).
We demonstrate how this layer abstracts schema complexity and aligns query intent with how CUs reason about their members.

% \vspace{0.1cm}
\textbf{Expressive querying for non-experts.}
We enable business users to ask complex analytical questions in natural language---without needing to know the underlying database schema.
Financial concepts like ``delinquency'', ``closure rate'', ``retention duration'', and ``longevity'' are rarely represented as explicit fields in the database.
Instead, they are often inferred from a combination of fields like balance, status, and dates.
We show how Tursio's prompt-based query rewriting capabilities allow users to express such complex concepts naturally, e.g., {\it ``Members with loans delinquent for more than $30$ days''}, {\it ``Develop a risk profile for accounts that display erratic behavior in transaction patterns''}, {\it ``Average retention duration for accounts by product category''},
without needing to know the underlying data representation.
We provide the contextual information about the database schema and available expressions, and use LLM to generate advanced SQL constructs including scalar functions, window functions, nested queries, and common table expressions.
This improved rewritten version maintains structural consistency while answering the questions more accurately.

% \vspace{0.1cm}
% \textbf{Understanding Domain Vocabulary.}
% This scenario is important because domain-specific vocabulary and terminology can vary significantly between organizations, even within the same industry.
% In credit unions, for example, different institutions may use different terms to describe similar concepts (such as ``closed'' vs. ``deactivated'' accounts, ``open'' vs. ``active'' accounts).
% We enable users to update domain vocabulary in the context graph using {\it enforcer rules}, enabling it to understand and map these terms correctly during query processing.
% This improves both the relevance and reliability of search results.

% \vspace{0.1cm}
\textbf{Resolving ambiguity with annotations.}
Real-world databases often contain ambiguities—such as columns with similar names but different meanings, or tables with overlapping terminology (e.g. column named {\it``close\_date''} appear in {\it LOAN}, {\it MEMBER\_ACCOUNT} and {\it CARD} tables).
When user asks to {\it ``List accounts which got closed last year''}, this can lead to LLM hallucination and thus incorrect query results.
We demonstrate how Tursio automatically surfaces such ambiguities and allows users to add annotations to the context graph to resolve them, ensuring accurate query interpretation.
User can add {\it prioritization rules} to indicate which table/column should be preferred when multiple options exist.
This not only improves the reliability of search results but also builds user trust in the system, which is critical for adoption in environments where data accuracy is paramount. We will let the audience play with different annotations and see how it affects the results.
% The feature to surface ambiguities will be ready by the time of camera-ready. 
% Given the ambiguities, how we resolve them using annotations has been implemented.

% \vspace{0.1cm}
\textbf{System validation and testing.}
The sensitivity of financial data, correctness of the results, and trust in the system are essential for adoption of a solution in credit unions.
We demonstrate Tursio's built in system-level testing and validation capabilities using features like {\it``Query awareness''}, {\it``Saving query''}, {\it``Providing feedback''}, {\it``Query history''}, and {\it``Insights''}.
Query awareness feature allows users to see how their queries are interpreted by the system before execution.
Users can provide positive or negative feedback on individual results. Negative feedback triggers a review: the system logs the query, generated SQL, and user correction. Administrators can then refine the context graph---e.g., adding annotations, adjusting join priorities, or adding enforcer rules---which take effect on subsequent queries after retraining the context graph.
Users can also bookmark interesting queries for future references. It can be used for reproducibility, consistent response, monitoring system performance, and auditing.
Bookmarked queries are visible to anyone having access to the same data source.
Query history feature enables users to track their interactions with the system.
We also provide insights on the query performance and usage patterns.
These features allow users to validate the system's accuracy, performance and track changes over time for safe adoption of the solution. The audience can test these features using their own queries and check the system reliability.

% \vspace{0.1cm}
% \textbf{Sharing and Collaboration.}
% We demonstrate how users can generate permalinks to share their queries and results, subject to access control policies.
% We enable users to export the result into PDF reports for offline sharing.

% \vspace{0.1cm}
% \textbf{Explaining Results to Stakeholders.}
% After executing the query, we generate a concise natural language explanation of the results tailored to the analytical intent (e.g., summarize vs. compare) of the query.
% For larger result sets, Tursio creates an agent to build a mini vector index over the results and generate explanations based on that.
% A user can optionally provide external context—such as recent economic or policy changes—which we can incorporates to provide more informed explanations.
% Stakeholders can then interactively ask follow-up questions and chat with the data to explore deeper insights.

% \vspace{0.1cm}
\textbf{Fine-grained access control.}
Finally, we illustrate how access control is enforced across different user roles. We support four roles: administrators, owners, users, and viewers.
Administrators and owners can manage system configurations and user permissions and have access to all the features. Users can only perform searches and view all the results.
A viewer role may have restricted access, allowing them to only query and view summary of the results by default.
We ensure that expressive search capabilities do not compromise data governance, auditability, or compliance requirements.
This role-based access control ensures that sensitive information is protected while still enabling effective data exploration.

% \vspace{0.1cm}
Overall, the above scenarios will allow the audience to experience Tursio's capabilities in addressing the unique challenges faced by credit unions --- making member insights accessible to their stakeholders reliably and securely.
\vspace{-0.2cm}

% \section{Conclusion}
% Credit unions face significant challenges in extracting member insights from complex, account-centric core banking schemas.
% Tursio addresses these challenges by providing a secure, context-aware database search solution that simplifies data access for users.
% Tursio offers three main advantages:
% (1) fully on-premises deployment, ensuring privacy, IT control, and predictable infrastructure-based costs;
% (2) automated construction of a context graph for deep contextual understanding, reducing ambiguity and eliminating manual sample question maintenance; and
% (3) accurate, context-aware search that generates well-formed query plans and delivers results in text, table, and visualization formats.

% Through realistic demonstration scenarios, we showcase Tursio's ability to handle member-centric queries, expressive querying, domain vocabulary understanding, ambiguity resolution, system validation, and access control.
% By bridging the gap between complex data structures and user intent, Tursio empowers credit unions to make informed decisions and enhance member services.

% \begin{acks}
% \end{acks}

\balance
\bibliographystyle{ACM-Reference-Format}
\bibliography{references}

% \appendix









\end{document}